\documentstyle[aps,prl,multicol]{revtex}    

\begin{document}

\draft

\title{Measurement of the vector analyzing power in elastic electron-proton 
scattering as a probe of double photon exchange amplitudes}
\author{S.P. Wells$^1$, T. Averett$^2$, D. Barkhuff$^3$, D.H. Beck$^4$, 
E.J. Beise$^5$, C. Benson$^4$, H. Breuer$^5$, R. Carr$^6$, S. Covrig$^6$, 
J. DelCorso$^{7}$\footnote{Present address: Department of Physics, 
College of William and Mary, Williamsburg, VA~23187}, 
G. Dodson$^3$\footnote{Present address: Spallation Neutron Source, 
Oak Ridge National Laboratory, Oak Ridge, TN 37830-8218},
C. Eppstein$^6$, M. Farkhondeh$^3$, B.W. Filippone$^6$, 
T. Forest$^{4}$\footnote{Present address: Department of Physics, 
Old Dominion University, Norfolk, VA~23529}, 
P. Frasier$^6$, R. Hasty$^4$, T.M. Ito$^6$, C. Jones$^6$, 
W. Korsch$^8$, S. Kowalski$^3$,
P. Lee$^6$, E. Maneva$^6$, K. McCarty$^6$, R.D. McKeown$^6$, 
J. Mikell$^4$, B. Mueller$^9$, P. Naik$^4$, M.L. Pitt$^7$, 
J. Ritter$^4$, V. Savu$^6$, D. T. Spayde$^5$, M. Sullivan$^6$, 
R. Tieulent$^5$, E. Tsentalovich$^3$, B. Yang$^3$, and T. Zwart$^3$}

\address{
$^1 Center~for~Applied~Physics~Studies,~Louisiana~Tech~University,
~Ruston,~LA~71272,~USA$\\
$^2 Department~of~Physics,~College~of~William~and~Mary,~Williamsburg,
~VA~23187,~USA$\\
$^3 Bates~Linear~Accelerator~Center,~Laboratory~for~Nuclear~Science~
and~Department~of~Physics,$\\ 
$Massachusetts~ Institute~of~Technology,~Cambridge,~MA~02139,~USA$\\
$^4 University~of~Illinois~at~Urbana-Champaign,~Urbana,~IL~61801,~USA$\\
$^5 University~of~Maryland,~College~Park,~MD~20742,~USA$\\
$^6 Kellogg~Radiation~Laboratory,~California~Institute~of~Technology,
~Pasadena,~CA~91125,~USA$\\
$^7 Department~of~Physics,~Virginia~Polytechnic~Institute~and~State
~University,$\\
$Blacksburg,~VA~24061,~USA$\\
$^8 Department~of~Physics~and~Astronomy,~University~of~Kentucky,
~Lexington,~KY~40506,~USA$\\
$^9 Physics~Division,~Argonne~National~Laboratory,
~Argonne,~IL~60439,~USA$\\}

\maketitle
\centerline{(SAMPLE Collaboration)}

\begin{abstract}

We report the first measurement of the vector analyzing
power in inclusive transversely polarized elastic electron-proton 
scattering at  $Q^2$~=~0.1 (GeV/c)$^2$ and large scattering 
angles.  This quantity should vanish in the single virtual photon 
exchange, plane wave impulse approximation for this reaction, 
and can therefore provide information on double photon 
exchange amplitudes for electromagnetic interactions with 
hadronic systems.  We find a non-zero value of 
$A$=-15.4$\pm$5.4~ppm.  No calculations of this observable 
for nuclei other than spin 0 have been carried out in these 
kinematics, and the calculation using the spin orbit interaction 
from a charged point nucleus of spin 0 cannot describe these data. 

\end{abstract}


\begin{multicols}{2}[]
\narrowtext

The recent development and refinement of experimental
methods for measurements of small (few parts per million,
or ppm) parity violating effects in polarized electron
scattering \cite{Spa99,HAP99,McK97} provides a new technique
for further studies of the electromagnetic
structure of hadrons and nuclei.  We have exploited these
methods for the first time to measure the small vector
analyzing power in elastic electron-proton scattering at
large scattering angles 
($130^{\circ} \le \theta ^e_{lab} \le 170^{\circ}$),
corresponding to $Q^2 =0.1$~(GeV/c)$^2$. This
parity conserving quantity is associated with transverse
electron polarization, in contrast to the parity violating
longitudinal (i.e. helicity dependent) asymmetry.
It has been previously noted \cite{Don86} that 
transverse polarization effects will be suppressed
by the relativistic boost factor 1/$\gamma$.
Nevertheless, as demonstrated in this Letter,
the development of the technology to measure
small parity violating asymmetries, along with the
ability to produce transversely polarized electron
beams at high energies,  now renders
these transverse polarization effects amenable
to measurement.

The vector analyzing power is a time reversal odd
observable which must vanish in first order
perturbation theory, and can only arise in leading order
from the interference of double photon exchange
(second order) and single photon exchange amplitudes.
Our observation of this quantity therefore demonstrates
the viability of a new technique to access physics 
associated with the absorption of two virtual photons 
by a hadronic system.  
Thus, the study of vector analyzing powers provides
another method to study double photon exchange
processes that is complementary to virtual compton
scattering (VCS).  VCS involves the coupling of
one virtual and one real photon to a hadronic system, 
but in practice includes problematic Bethe-Heitler
amplitudes associated with radiation of a real
photon from the electron. Nonetheless, there is
presently a great deal of interest in the exploitation
of VCS \cite{Gui98} to further probe the electromagnetic
structure of hadrons and nuclei, and the vector 
analyzing power described here potentially offers an attractive 
alternative to access double photon exchange amplitudes.

Using the apparatus for the SAMPLE experiment 
\cite{Spa99,Mue97}, a high statistics measurement of the 
parity violating asymmetry in inclusive elastic 
$p(\vec e, e^{\prime})$ scattering at the  MIT/Bates 
Linear Accelerator Center, we have made measurements
of the asymmetry in the elastic scattering of 200~MeV 
transversely polarized electrons from the proton at 
backward scattering angles.
This represents the first measurement of a vector 
analyzing power in polarized electron scattering 
from the proton at this high  a momentum transfer.

The vector analyzing power in electron-nucleus scattering
results in a spin-dependent asymmetry,  which can, 
for example, be generated by the interaction of the electron 
spin with  the magnetic field seen by the electron in its 
rest frame \cite{Gay92}.  This spin-dependence in the scattering 
cross section $\sigma (\theta )$, can be written as \cite{Kes85,Ohl72}
\begin{equation}
\sigma (\theta ) = \sigma _0 (\theta )[1+A(\theta ) {\bf P \cdot \hat n} ],
\end{equation}
where $\sigma _0 (\theta )$ is the spin-averaged scattering 
cross section, $A(\theta )$ is the vector analyzing power for the 
reaction, and ${\bf P}$ is the incident electron polarization vector 
(which is proportional to the spin vector operator ${\bf S}$).  
The unit vector 
${\bf \hat n}$ is normal to the scattering plane, and is defined through 
${\bf \hat n} \equiv 
({{{\bf k} \times {\bf k ^{\prime}}}})$/$\vert {\bf k} \times 
{\bf k ^{\prime}} \vert$,
where ${\bf k}$ and ${\bf k ^{\prime}}$ are wave vectors for the 
incident and scattered electrons, respectively.  
The scattering angle $\theta$ is found through 
$\cos \theta = ({\bf k} \cdot {\bf k ^{\prime}})$/$\vert {\bf k} \vert 
\vert {\bf k ^{\prime}} \vert$, and, in the Madison convention, 
is positive for the electron scattering toward the same direction 
as the transverse component of ${\bf k^{\prime}}$.
The beam polarization ${\bf P}$ can be expressed in terms of the 
number of beam electrons with spins parallel ($m_s$=+1/2) and 
antiparallel ($m_s$=-1/2) to ${\bf \hat n}$, so that the 
measured asymmetry at a given scattering angle $\epsilon (\theta )$, 
is defined through
\begin{equation}
\epsilon (\theta ) = 
{{\sigma _{\uparrow} (\theta )- \sigma _{\downarrow} (\theta )}
\over{\sigma _{\uparrow} (\theta )+ \sigma _{\downarrow} (\theta )}} 
= A(\theta )\langle P \rangle ,
\end{equation}
where $\sigma _{\uparrow ,\downarrow}(\theta )$ is the differential
cross section for $m_s$ = +1/2 and -1/2, respectively.
Thus, with knowledge of the magnitude of the incident beam polarization 
$\langle P \rangle$, measurement of $\epsilon (\theta )$ can yield a 
determination of the vector analyzing power $A(\theta )$, 
which contains the underlying physics of 
the electron-nucleus interaction.

The formalism and conventions reviewed here have been well established, 
and used extensively for ``Mott'' polarimeters which measure 
electron beam polarizations at low incident beam energies 
($\sim$ 100~keV) \cite{Gay92},  where it is valid to assume the nucleus 
is simply a point charge.  The ``Mott'' asymmetry for 
transversely polarized electrons scattering from a point nucleus of 
charge $Ze$ and spin 0 is calculated as \cite{Mot32},
\begin{equation}
A_{Mott} = -{{4Z\alpha \beta}\over{\gamma}}
({{\csc \theta \ln (\csc {{\theta}\over{2}})}
\over{\csc ^4 {{\theta}\over{2}}-\beta ^2 \csc ^2 {{\theta}\over{2}}}}),
\end{equation}
where $\alpha$ is the electromagnetic fine structure constant, 
$\beta$ and $\gamma$ are the usual relativistic kinematic quantities, 
and $\theta$ is the laboratory scattering angle defined above. 
The analyzing powers calculated via Eq. (3) for low energies
using very high $Z$ targets are much larger than
the vector analyzing power for electron proton
scattering reported here, and this is commonly 
exploited as a means to measure the
polarization of low energy electron beams.  Such
measurements, however, are not sensitive to the
internal structure or spin of the hadronic system.
More recently, some level of nuclear structure has
been taken into account in calculations of analyzing powers 
for high energy elastic scattering of transversely polarized 
electrons from heavy spin 0 nuclei at forward electron scattering 
angles, performed in the eikonal expansion and using finite charge 
densities for the nuclei \cite{Kro89}. 
In these calculations, the non-zero analyzing powers were
generated through the distortion of the electron waves in the
Coulomb potential of the nuclear targets, providing the needed
extension beyond single photon exchange to the distorted wave impulse 
approximation. To date, however, no such calculation 
exists for the scattering of transversely polarized electrons from 
nuclear targets of any spin other than 0 at any energy. 
In the SAMPLE kinematics, the electron energy of 200~MeV is much 
larger than the energies used for Mott polarimetry, the proton target 
has the smallest possible $Z$ so that Coulomb effects are at a minimum, 
and the electrons are scattered at large angles where 
magnetic effects are important.  These facts, along with the 
spin 1/2 nature of the proton, imply that our measurement of 
the vector analyzing power will be sensitive to non-trivial 
electromagnetic structure of the proton not taken into account in 
previous theoretical treatments.

The data we report are the result of an experiment  performed at the 
MIT/Bates Linear Accelerator Center with a 200~MeV polarized electron 
beam of average current 40~$\mu$A incident on a 40~cm liquid
hydrogen target \cite{Be96}.  The  scattered electrons 
were detected in a large solid angle ($\sim$~1.5~sr), axially symmetric
air \v Cerenkov detector consisting of 10 mirrors, each shaped to focus the 
\v Cerenkov light onto one of ten shielded photomultiplier tubes. This
combination of large solid angle and high luminosity allow measurements 
of small asymmetries in a relatively short period of time.  The
data presented here were acquired in just two days of running under these 
conditions. Properties of the detector signals and  beam have been described 
in detail in Ref.'s \cite{Spa99} and \cite{Mue97}, along with the method of 
asymmetry extraction and correction.  Thus, here we report only the differences 
between the experimental running conditions for longitudinally polarized
beams as used for parity violation measurements, and the transversely 
polarized beam used for the vector analyzing power measurements. 
The systematic errors associated with the asymmetries from each of
the individual mirrors are the same for these measurements as for
those in Ref.'s  \cite{Spa99} and \cite{Mue97}, totaling 0.7~ppm, and
are negligible compared with the overall statistical error of 5.4~ppm
obtained for these measurements. 

The polarized laser light used on the bulk GaAs source crystal 
produces electron beams with longitudinal polarization, consequently
significant spin manipulation was required to orient the beam polarization 
transversely.  This was achieved with a Wien filter, which contains electric 
and magnetic fields oriented perpendicular to each other and to the beam 
direction, and a set of  beam solenoids.  The Wien filter was positioned  
immediately downstream of the source anode, and was used to precess 
the electron spin away from the beam direction ($\sim$ 90$^{\circ}$ for 
these measurements).  The beam solenoids were positioned near the first 
accelerating cavity in the beam line, and precessed the resulting transverse 
components of the beam polarization. The combination of 
these beam line elements allowed the polarization direction to be 
chosen arbitrarily, and each element was calibrated such that the 
polarization direction is determined to $\pm$~2$^{\circ}$ \cite{PittTBP}. 

For the measurements reported here, 
two orthogonal transverse beam polarizations were used during two  
running periods: one with the polarization directed to beam right 
(which we denote $\Phi$=0), and one with the polarization
pointing up ($\Phi$=90).  The magnitude of the beam polarization was 
measured with a Moller apparatus positioned on the beam line, and averaged 
36.3$\pm$1.8\% during these measurements. Finally, to minimize 
false asymmetries and test for systematic errors, the electron beam 
polarization was manually reversed relative to all electronic signals, 
for both $\Phi$=0 and $\Phi$=90 running, with the insertion 
of a $\lambda$/2 plate in the laser beam. Thus, four separate sets of 
measurements were made: $\Phi$=0, $\lambda$/2 IN and OUT, and
$\Phi$=90, $\lambda$/2 IN and OUT.

The elastic scattering transverse asymmetry was determined 
for each of the 10 individual mirrors in each running configuration
after correction for all effects, including beam polarization, background 
dilution, and radiative effects, as described 
in Ref.'s \cite{Spa99} and \cite{Mue97}. Although the geometry of this 
detector allowed for combining the asymmetries from  individual 
mirrors positioned on opposite sides of the incident beam (via Eq. (2) and 
imposing the rotational invariance criterion $A(\theta ) = -A(-\theta )$ 
\cite{Ohl72}), we chose an alternative form of analysis
wherein the full statistical information contained in the data set could
be used to extract the vector analyzing power.
Because the individual mirrors were positioned at varying 
azimuthal angles $\phi$ relative to the polarization direction, 
the asymmetries measured in the mirrors should follow a sinusoidal 
dependence in this angle.  The sinusoidal dependence in the azimuthal 
angle $\phi$ is seen by rewriting Eq. (2) as
\begin{equation}
\epsilon (\theta ,\phi) = A(\theta ) P \sin (\phi +\delta ),
\end{equation}
where $\phi$ measures the angle of the polarization vector in
the plane transverse to the beam direction, and the 
phase $\delta$ takes into account the direction of ${\bf P}$
relative to ${\bf \hat n}$. Table 1 summarizes the polar ($\theta$) and 
azimuthal ($\phi$) angles at the center of each individual mirror 
within the SAMPLE detector. As seen in Table 1, mirrors 4 and 5 have 
the same azimuthal angle relative to the polarization direction, but 
different polar angles relative to the incident beam direction.  
A separate analysis, however, indicated that the polar angle 
dependence to the asymmetry was negligible, allowing us to combine 
the asymmetries from these two mirrors (similarly for mirrors
6 and 7) into one asymmetry at the same azimuthal angle $\phi$. 

The  data set for each $\Phi$ and $\lambda$/2 running configuration 
therefore consists of eight data points at varying $\phi$ values, to which 
we perform a $\chi ^2$ minimization to a two parameter function via
\begin{equation}
\chi ^2 _{d.o.f.} = {{1}\over{6}} \sum _{i=1}^8 [A ^{meas}_i - 
(a\sin \phi _i + b\cos \phi _i )]^2 /[\delta A^{meas}_i]^2  ,
\end{equation}
which is linear in the coefficients $a$ and $b$. 
Here $A^{meas}_i$ is the measured asymmetry at each azimuthal 
angle $\phi _i$, corrected for all effects \cite{Spa99,Mue97}
including beam polarization normalization (as suggested in Eq. (4)). 
The coefficients $a$ and $b$ can then be converted into 
an amplitude and phase, i.e.,
\begin{equation}
A_{fit}=A\sin (\phi +\delta )
\end{equation}
as in Eq. (4), where the amplitude $A$ gives the magnitude of  the 
vector analyzing power, and the phase $\delta$ verifies the 
direction of the beam polarization and determines the overall sign of 
the analyzing power.

The sinusoidal dependence just discussed is illustrated in Fig. 1, 
where the combined data for $\Phi=0$ and $\Phi=90$ 
are shown as a function of azimuthal angle, along 
with the best fit to the data according to the procedure outlined above. 
Here we have defined $\phi=0$ to be at  beam left, 
and have taken into account the 90$^{\circ}$ phase difference between 
the $\Phi=0$ and $\Phi=90$ polarization directions.  For these combined 
data, the overall $\chi ^2$ per degree of freedom for the best fit was 
found to be 0.9, providing a 50\% confidence level that the data follow 
this dependence \cite{PDB}.   This should be compared, however, with 
the $\chi ^2$ per degree of freedom of 2.1 for a
fit to $A=0$, which has a corresponding confidence level of 4\% that the
data are consistent with $A=0$. Even if we allow an overall offset to
a constant dependence, we find an average of $A=3.5 \pm 3.7$~ppm, 
with a $\chi ^2$ per degree of freedom of 1.9, and a corresponding 
confidence level of 7\%.

In Table 2 we summarize our results using this analysis procedure for 
the four independent running conditions.  Note that the deduced 
magnitudes are all consistent within experimental errors,
and the deduced phase changes by 180$^{\circ}$ upon the insertion 
or removal of the $\lambda$/2 plate as expected, and by 90$^{\circ}$ 
from one $\Phi$ running configuration to the other. Combining these  
four independent measurements, we  quote our final result:   a vector 
analyzing power for elastic electron-proton scattering of 
-15.4$\pm$5.4~ppm at the average electron laboratory scattering 
angle of 146.1$^{\circ}$, corresponding to $Q^2$~=~0.1~(GeV/c)$^2$. 
To demonstrate the precision to which this quantity
has been determined relative to the original derivation of 
Mott \cite{Mot32}, we plot this data point in Fig. 2 along with the 
prediction of Eq. (3) for a point nucleus of charge $Z=1$ and spin 0 
as a function of electron laboratory scattering angle, covering the 
angular range accepted by the SAMPLE detector.  

The data reported here represent the first measurement of a vector
analyzing power in polarized electron scattering at this high a 
momentum transfer.   
Our observation of this quantity demonstrates the viability of a new
technique to access physics associated with double photon exchange,
which may address some of the same physics issues as virtual
compton scattering measurements. We have also made measurements 
of the vector analyzing power in inclusive quasielastic electron-deuteron 
scattering, the results of which will be reported in a future letter.  
Further parity violation measurements at higher $Q^2$ values 
are planned \cite{TJNAF} from both hydrogen and deuterium targets, 
where high statistics transverse asymmetry data will also be 
taken. Thus, we hope that the results reported in this Letter will
motivate theoretical calculations of vector analyzing powers
in polarized electron scattering for hadronic systems with
$S \neq 0$ and non-trivial electromagnetic structure, which will
be necessary to interpret such measurements.

The efforts of the staff at MIT/Bates to provide high quality beam required 
for these measurements, and useful conversations with T.W. Donnelly, are 
gratefully acknowledged.

This work was supported by NSF grants PHY-9870278 (Louisiana Tech), 
PHY-9420470 (Caltech), PHY-9420787 (Illinois), PHY-9457906/PHY-9229690 
(Maryland), PHY-9733773 (VPI) and DOE cooperative agreement 
DE-FC02-94ER40818 (MIT/Bates) and contract W-31-109-ENG-38 (ANL).

\begin{table}
\begin{center}
\begin{tabular}{ccc}
\multicolumn{1}{c}{Mirror} {\vline}&
\multicolumn{1}{c}{$\theta$ (deg.)} {\vline}&
\multicolumn{1}{c}{$\phi$ (deg.)}\\
\hline\hline
  1  &  146  &  135  \\
  2  &  154  &  90  \\
  3  &  146  &  45  \\
  4  &  138  &  180  \\
  5  &  161  &  180  \\
  6  &  161  &  0  \\
  7  &  138  &  0  \\
  8  &  146  &  225  \\
  9  &  154  &  270  \\
  10  &  146  &  315  \\
\end{tabular}
\end{center}
\caption{Polar ($\theta$) and azimuthal ($\phi$) angles of each individual
mirror within the SAMPLE detector.}
\label{tabl:tab1}
\end{table}

\begin{table}
\begin{center}
\begin{tabular}{ccccc}
\multicolumn{1}{c}{$\Phi$} {\vline}&
\multicolumn{1}{c}{$\lambda$/2} {\vline}&
\multicolumn{1}{c}{$A$ (ppm)} {\vline}&
\multicolumn{1}{c}{$\delta$ (deg.)} {\vline}&
\multicolumn{1}{c}{$\chi ^2 _{d.o.f.}$}\\
\hline\hline
  0        & IN & 12.9$\pm$9.8 & 173.8$\pm$39.5 & 1.30\\
  0        & OUT & 13.8$\pm$9.9 & 16.9$\pm$39.5& 1.50 \\
  90        & IN & 18.4$\pm$11.8 & -84.1$\pm$39.8& 0.30 \\
  90        & OUT & 18.1$\pm$11.7 & 127.2$\pm$38.0& 2.07\\
\end{tabular}
\end{center}
\caption{Results of the fitting procedure described in the text.}
\label{tabl:tab2}
\end{table}

\begin{figure}
\caption{Plot of the measured asymmetry, corrected for all
effects including beam polarization, background dilution, and 
radiative effects, as a function of azimuthal scattering angle 
$\phi$ for the combined data of all four running configurations 
as described in the text. The curve represents 
the best fit to the data according to Eq. (5).}
\end{figure}

\begin{figure}
\caption{Plot of the measured vector analyzing power in elastic
electron-proton scattering in the SAMPLE
kinematics, along with the prediction of the original Mott 
derivation [10], given in Eq. (3).}
\end{figure}

\end{multicols}

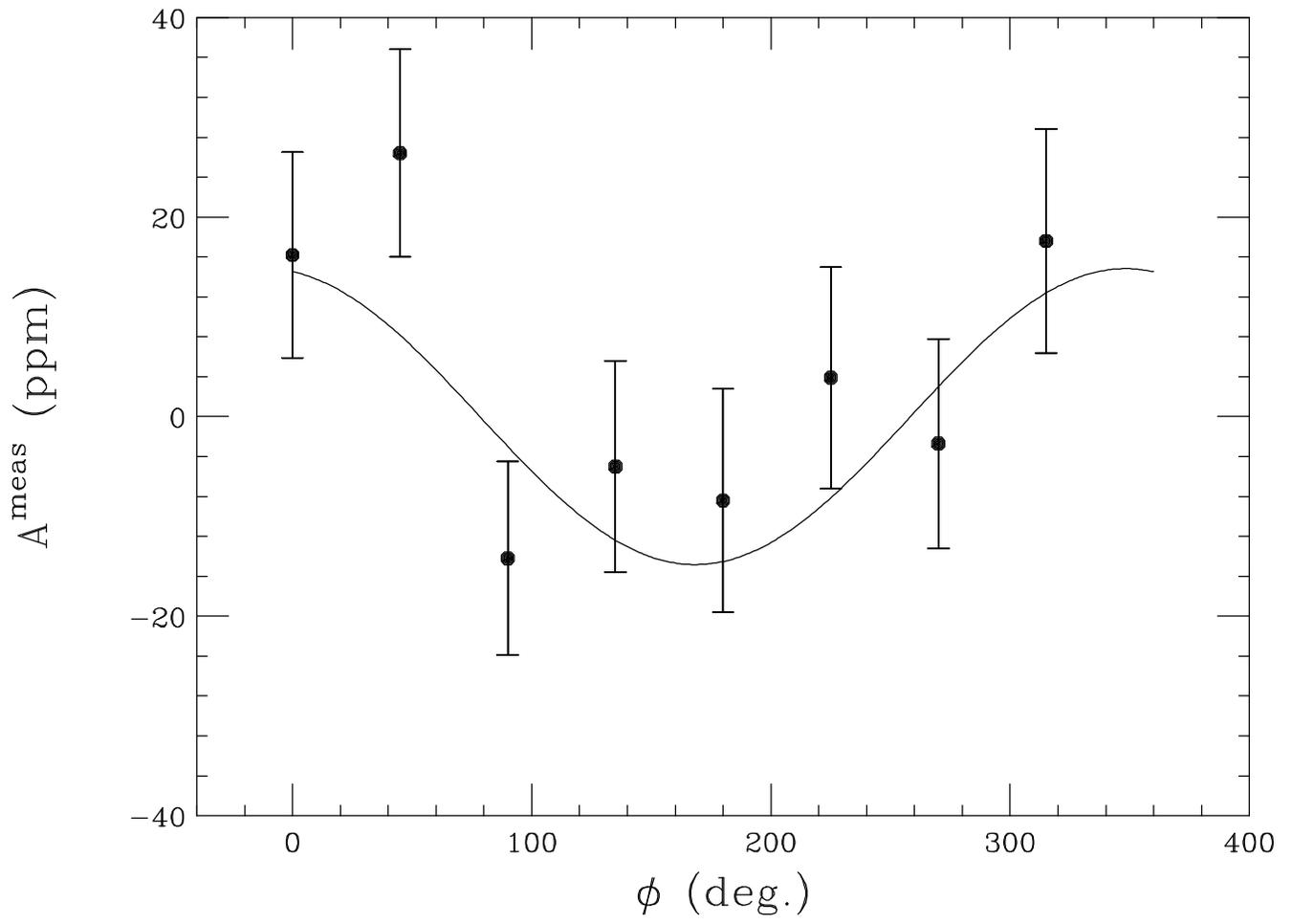
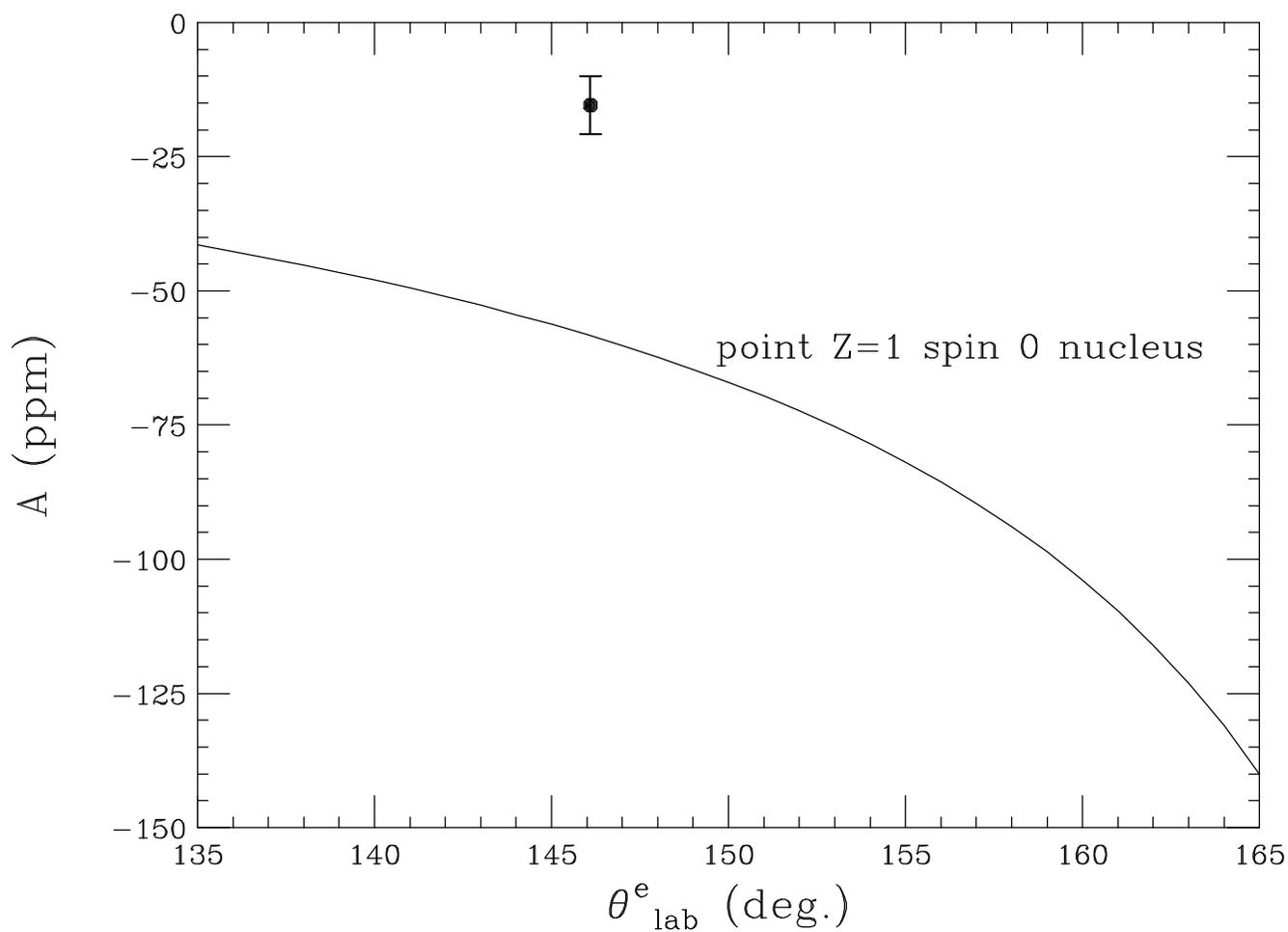
\end{document}